\documentclass[%
 reprint,
 superscriptaddress, 
 amsmath,amssymb,
 aps,
 prl,
]{revtex4-2}

\usepackage{xcolor}
\usepackage{graphicx}
\usepackage{dcolumn}
\usepackage{bm}

\newcommand\mPas{\textrm{mPa}\cdot\textrm{s}}

\begin{document}

\preprint{APS/123-QED}

\title{Rheofluidics: frequency-dependent rheology of single drops}

\author{Matteo Milani}
\thanks{These authors contributed equally to this work.}
\affiliation{PMMH, ESPCI, Universit\'e PSL, CNRS, Sorbonne Universit\'e, Universit\'e Paris Cit\'e, 75005 Paris, France}
\affiliation{John A. Paulson School of Engineering and Applied Sciences, Harvard University, Cambridge, Massachusetts 02138, USA}

\author{Wenyun Wang}
\thanks{These authors contributed equally to this work.}
\affiliation{John A. Paulson School of Engineering and Applied Sciences, Harvard University, Cambridge, Massachusetts 02138, USA}

\author{Lorenzo Russotto}
\affiliation{Department of Physics, Harvard University, Cambridge, Massachusetts 02138, USA}

\author{Weiyu Zong}
\affiliation{John A. Paulson School of Engineering and Applied Sciences, Harvard University, Cambridge, Massachusetts 02138, USA}

\author{Kevin Jahnke}
\affiliation{John A. Paulson School of Engineering and Applied Sciences, Harvard University, Cambridge, Massachusetts 02138, USA}
\altaffiliation{present address: Max Planck Institute for Medical Research}

\author{David A. Weitz}
\affiliation{John A. Paulson School of Engineering and Applied Sciences, Harvard University, Cambridge, Massachusetts 02138, USA}
\affiliation{Department of Physics, Harvard University, Cambridge, Massachusetts 02138, USA}

\author{Stefano Aime}
\email{stefano.aime@espci.fr}
\affiliation{John A. Paulson School of Engineering and Applied Sciences, Harvard University, Cambridge, Massachusetts 02138, USA}
\affiliation{Molecular, Macromolecular Chemistry, and Materials, ESPCI Paris, 75005 Paris, France.}

\date{\today}

\begin{abstract}
We present Rheofluidics, a microfluidic technique that measures the frequency-dependent rheology of individual micron-scale objects. We apply oscillatory hydrodynamic stresses by flowing them through channels with modulated constrictions, and measure their deformation.
Unlike bulk rheology, which measures collective properties, Rheofluidics provides heretofore unattainable measurements of individual particles.
We apply Rheofluidics to discover frequency-dependent surface tension of surfactants, very high-frequency viscoelasticity of microgels and unexpected frequency-dependent bending modulus of vesicles.
\end{abstract}

\maketitle

Rheology provides a powerful means to characterize the mechanical response of materials by measuring their viscoelastic moduli as a function of frequency. In bulk systems, this is routinely done using commercial rheometers that measure viscoelastic moduli by applying oscillatory strain or stress, and measuring the corresponding response~\cite{macoskoRheologyPrinciplesMeasurements1994}. 
Viscoelastic properties are also important to understand the behavior of microscopic objects such as bubbles and droplets \cite{stoneDynamicsDropDeformation1994,bochnerdearaujoDropletCoalescenceSpontaneous2017}, capsules and vesicles \cite{feryMechanicalPropertiesMicro2007,dimovaGiantVesiclesTheir2019,morshedMechanicalCharacterizationVesicles2020}, or even cells and organelles \cite{shenLiquidtosolidTransitionFUS2023,urbanskaSinglecellMechanicalPhenotype2017}. Rheometers can measure bulk samples, which can only be packings of these objects; when applied to bulk emulsions, foams and tissue, rheology offers detailed insights into the collective behavior of droplets, bubbles and cells, but remains largely insensitive to the viscoelasticity of the individual components themselves. 
Measurement of the mechanical properties of such microscopic objects remains a long-standing challenge~\cite{harvey_properties_1938}. Existing techniques can provide static measurements based on compression~\cite{cole_surface_1932,liuCompressiveDeformationSingle1996}, indentation~\cite{mcconnaughey_cell_1980,tao_measuring_1992}, aspiration~\cite{mitchison_mechanical_1954,rand_mechanical_1964,hochmuthMicropipetteAspirationLiving2000}, or stretching~\cite{taylorFormationEmulsionsDefinable1934,henonNewDeterminationShear1999,guckOpticalStretcherNovel2001}. 
Alternative approaches using microfluidic devices enable high-speed deformation measurements of cells, bubbles and droplets flowing through narrow channels \cite{mietkeExtractingCellStiffness2015, dapolitoMeasuringInterfacialTension2018}, junctions \cite{ulloaEffectConfinementDeformation2014,dengInertialMicrofluidicCell2017} or constrictions \cite{mulliganEffectConfinementinducedShear2011,martinInterfacialRheologyMicrofluidics2011,brosseauMicrofluidicDynamicInterfacial2014}, while precise frequency-dependent viscoelastic measurement of the moduli of these objects remains elusive~\cite{tregouetTransientDeformationDroplet2018}. This hinders a more complete understanding of their properties and their individual contribution to bulk rheology.

Here, we introduce Rheofluidics, a microfluidic technique that enables measurement of the frequency-dependent moduli of individual objects, including drops, microgel particles, and vesicles. By using channels with periodically varying constrictions, we apply oscillatory hydrodynamic stresses and measure the time-dependent deformation of microscopic objects in the flow. This enables the measurement of frequency-resolved viscoelastic moduli, analogous to stress-controlled rheology of bulk materials.
Because of the low inertia and fast flow rates typical of microfluidics, Rheofluidics enables very high-frequency behavior to be probed. We illustrate the utility of Rheofluidics by investigating three very different systems. We investigate the frequency-dependent surface tension of surfactant-stabilized emulsion droplets, and demonstrate that we are able to explore the absorption time of surfactant molecules at the interface. 
We also investigate hydrogel beads, and show that we are able to determine their viscoelastic moduli in frequency ranges inaccessible to rheometers.
Finally, we measure the mechanics of lipid vesicles and show that their  deformation is dominated by the nonlinear bending modulus of the membrane, which we characterize as a function of vesicle composition and frequency.
These measurements demonstrate how Rheofluidics measures the rheology of a wide variety of soft matter systems at the single particle level.

The key concept of Rheofluidics is the use of a microfluidic channel where the flow of the continuous fluid phase applies a planar extensional stress with a well-defined temporal evolution on individual objects suspended in the fluid. 
To impose an oscillatory profile, $\sigma_{ext}(t)=\sigma_0 \sin(\omega t)$, we design a series of constrictions in a channel with rectangular cross-section with a thickness $h=60~\mu m$ and width $L(x)$, which varies periodically along the channel axis. 
The channel shape forces the velocity flow field, $\vec{v}$, to repeatedly converge and diverge, resulting in oscillations in the extensional stress, $\sigma_{ext}=\eta \dot\varepsilon$, where $\eta$ is the viscosity of the fluid and $\dot\varepsilon=(\partial_x v_x-\partial_y v_y)/2$ is the extensional strain rate. Mass conservation for the incompressible continuous phase in this planar flow field requires that $\partial_xv_x+\partial_yv_y=0$, such that $\dot\varepsilon=\partial_xv_x$ is set by the change in flow speed due to the change in $L(x)$. 
We then determine the channel shape by inverting the relationship between $L(x)$ and $\sigma_{ext}(t)$, computed in the reference frame of the continuous phase. To this end, we express the flow speed in the center of the channel as $v_x=q/L$, where $q$ is an effective planar flow rate. For a straight channel, Poiseuille laminar flow yields $q=\beta Q/h$, where $Q$ is the volumetric flow rate, and $\beta$ is a coefficient that varies weakly between $2/3$ and 1 as the channel aspect ratio, $L/h$, decreases~\cite{SM}. Here, we neglect the spatial variation of $q$ arising from that of $L(x)/h$, and we obtain $\sigma_{ext} = \eta \partial_x v_x =- \eta q L^\prime/L^2$, where $L^\prime \equiv dL/dx$. To prescribe $L(x)$ such that $\sigma_{ext}$ is sinusoidal \textit{in time}, we express time as a function of position, in the reference frame of the flowing fluid: $t(x)=\int_0^x ds/v_x(s)$. 
We use this result in the expression of $\sigma_{ext}(t)$ to obtain an integro-differential equation for $L(x)$: 

\begin{equation}
\frac{dL}{dx}=L^2 \frac{\sigma_0}{q\eta} \sin\left[\frac{\omega}{q} \int_0^x L(s) ds\right]
\label{eq:chshape}
\end{equation}

\begin{figure}[ht]
\includegraphics[width=\columnwidth]{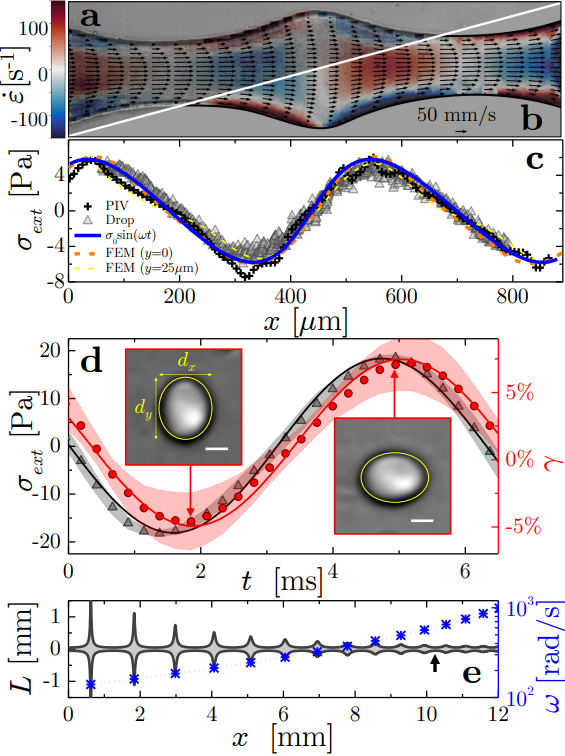}
\caption{\label{fig.1}\textbf{Channel design} (a) Microscopy image of a channel section obtained for $L_0=100~\mu$m, $\tilde{\sigma}L_0^2=0.23$ and $\tilde{\omega}L_0^2=0.46$. Black arrows: flow speed. Color scale: extensional rate. (b) FEM simulation results. (c) extensional stress from PIV (black crosses), FEM along the channel axis and 25~$\mu$m away (orange and yellow dashed lines) and droplet speed (gray triangles). Blue line: prescribed sinusoidal stress \textit{in time}. (d) Gray triangles, left axis: extensional stress for $\omega=1000$~rad/s. Red circles, right axis: droplet deformation. Insets: snapshots of droplets ar maximum and minimum deformation. Scale bar is 10~$\mu$m. (e) Black lines, left axis: \textit{chirp} Rheofluidic channel shape. Blue symbols, right axis: frequency for different channel sections.}
\end{figure}

This equation has two parameters, $\tilde\sigma = \sigma_0/q\eta$ and $\tilde\omega = \omega/q$, which are the rescaled amplitude and frequency of stress oscillations. 
We solve it numerically starting from $L(0)=L_0$ to obtain $L(x)$, and we use this solution to fabricate the microfluidic channels using soft lithography. A typical example of the channel shape exhibits a periodic series of constrictions of width $L_0$, separated by wider channel sections where the flow expands and contracts to impose the oscillatory $\sigma_{ext}$, as shown in Fig.~\ref{fig.1}a. 

To validate the assumptions used to derive $L(x)$, we simulate pressure-driven laminar flow in the full 3D geometry by solving Stokes' equation using the finite element method (FEM). We obtain a prediction for the velocity flow field, $\vec{v}$, and the extensional flow rate, $\dot\varepsilon$, in the midplane of the channel, shown as arrows and color maps in Fig.~\ref{fig.1}b. We compare them to experimental results obtained by Particle Image Velocimetry (PIV) analysis of a suspension of 500nm tracer particles in a 1:3 water:glycerol mixture with $\eta=58~\mPas$ flowing through the channel. 
Experimental values of $\vec{v}$ and $\dot\varepsilon$ (Fig.~\ref{fig.1}a) are in very good agreement with numerical simulations (Fig.~\ref{fig.1}b). 
Along the channel axis, we obtain oscillations of $\sigma_{ext}=\eta\dot\varepsilon$ that closely match the profile prescribed by the channel design, as shown by the blue solid line in Fig.~\ref{fig.1}c. We note that achieving the sinusoidal variation of stress in time requires a slightly non-sinusoidal stress in space, reflecting the inhomogeneous flow speed.
Away from the channel axis, $\sigma_{ext}$ deviates from the prescribed sinusoidal profile, as shown by the fading color shades in Fig.~\ref{fig.1}a-b. To ensure that the droplets are subject to a nearly homogeneous $\sigma_{ext}$, we restrict the analysis to droplets small enough that the maximum spatial variation of $\sigma_{ext}$ is of the order of the experimental uncertainty, around 20\%. Based on FEM simulations, we estimate that this corresponds to droplets of diameter smaller than 50~$\mu$m for this channel, which is half the minimum channel width. We expect that such small droplets will flow at a speed close to that of the outer fluid~\cite{happel_low_1983}. Under this working hypothesis, they will be subject to the prescribed oscillatory extensional stress.

To test this, we produce droplets of silicon oil stabilized by polysorbate 80 (Tween80) in 1:3 water:glycerol, and we flow them in the Rheofluidics channel, imaging them with an optical microscope equipped with a fast camera. The droplets have a radius $r_0\approx 9~\mu$m. 
We reconstruct both the flow field of the continuous phase, using PIV, and the speed of the droplets' center of mass, $v_d$, measured by tracking their position. We check that $v_d$ nearly matches the flow speed of the surrounding fluid. This provides us with an easier method of measuring the extensional stress applied to the droplet, $\sigma_{ext}\approx \eta \partial v_d/\partial x$. This result is equivalent to that obtained by PIV, as shown by comparing black crosses and gray triangles in Fig.~\ref{fig.1}c. In the following, we will extract $\sigma_{ext}$ from the droplet speed, as this method does not require seeding the continuous phase with tracer particles, and because it is more robust under larger applied flow rates, thus larger stresses. 
By increasing the flow rate, we obtain a large oscillatory frequency, $\omega=1000$~rad/s, with droplets covering the entire field of view in less than 10~ms. Analyzing their trajectory, we find that $\sigma_{ext}(t)$ is very well described by a sinusoidal oscillation with amplitude $\sigma_0=18$~Pa, as shown by the gray triangles in Fig.~\ref{fig.1}d. As the droplets flow through the channel constrictions, we find that their shape is periodically deformed, as shown in the insets of Fig.~\ref{fig.1}d. To quantify this effect, we measure their size along the channel axis and perpendicular to it, $d_x$ and $d_y$, and we define the droplet deformation, $\gamma=(d_x-d_y)/(d_x+d_y)$. This definition of $\gamma$, first employed by Taylor~\cite{taylorFormationEmulsionsDefinable1934}, captures the elliptical deformation of a droplet in an extensional flow field \cite{coxDeformationDropGeneral1969}. It also describes the leading-order expansion of a generic droplet deformation~\cite{mietke2015extracting}. Here, we verify that the shape of the droplet in the image plane is indeed elliptical to within experimental uncertainty~\cite{SM}. This confirms that such small droplets are subject to a homogeneous extensional stress field. 
By measuring $\gamma$ as a function of time, we find that it oscillates with the same frequency as $\sigma_{ext}$, and amplitude $\gamma_0=6\%$. Surprisingly, we find that the two signals have a measurable phase lag, $\varphi=0.3$~rad, which allows us to separate the in-phase and the out-of-phase components of the droplet response to the applied oscillation, yielding elastic and viscous moduli, $G^\prime$ and $G^{\prime\prime}$, as in a rheology experiment \cite{macoskoRheologyPrinciplesMeasurements1994}. 
For the oil droplet measured in Fig.~\ref{fig.1}d, we find that $G^\prime=300$~Pa and $G^{\prime\prime}=100$~Pa, indicating a predominantly elastic mechanical response: for liquid droplets with a radius $r_0$ at rest, $G^\prime$ is set by the Laplace pressure, and can be used to measure the surface tension, $\Gamma$. Studying droplets deformed under a constant extensional stress, Taylor~\cite{taylorFormationEmulsionsDefinable1934} first derived that  $\gamma=2\sigma_{ext}r_0/\Gamma$, an expression also used in microfluidic interfacial tensiometry~\cite{hudsonMicrofluidicInterfacialTensiometry2005,brosseauMicrofluidicDynamicInterfacial2014,chenSizeDependentDroplet2020}. 
Using this relationship, we obtain $\Gamma=2G^\prime r_0=6$~mN/m, slightly larger than that measured by pendant drop tensiometry, $\Gamma_{pd}=4$~mN/m. 

To investigate the origin of this discrepancy, we exploit the flexibility of Rheofluidics, and extend this measurement to the full spectrum of accessible frequencies. This can be done, for a given channel shape, by changing the flow rate, $q$. If $q$ is decreased, the droplets flow through the constrictions in a longer time, resulting in a lower $\omega$, as the channel shape sets $\tilde\omega=\omega/q$. However, doing so decreases proportionally the amplitude of the stress oscillations, $\sigma_0$, as $\tilde\sigma=\sigma_0/q\eta$ is set by the channel shape.
To control $\sigma_0$ and $\omega$ independently, and to increase the range of accessible $\omega$, we modify the Rheofluidic channel by gradually changing the shape of consecutive constrictions to obtain a \textit{chirp} stress signal~\cite{bouzidComputingLinearViscoelastic2018}, as shown in Fig.~\ref{fig.1}e. Analyzing the droplet oscillation in different channel sections, for a fixed $q$, we measure viscoelastic moduli as a function of $\omega$, as shown by blue symbols in Fig.~\ref{fig.1}e. 
Because we can image only a few constrictions, we reconstruct the frequency spectrum by taking several videos of different channel sections. For each section, we analyze 10-100 droplets of a given size to obtain average values of $G^\prime(\omega)$ and $G^{\prime\prime}(\omega)$.

\begin{figure}[ht]
\includegraphics[width=\columnwidth]{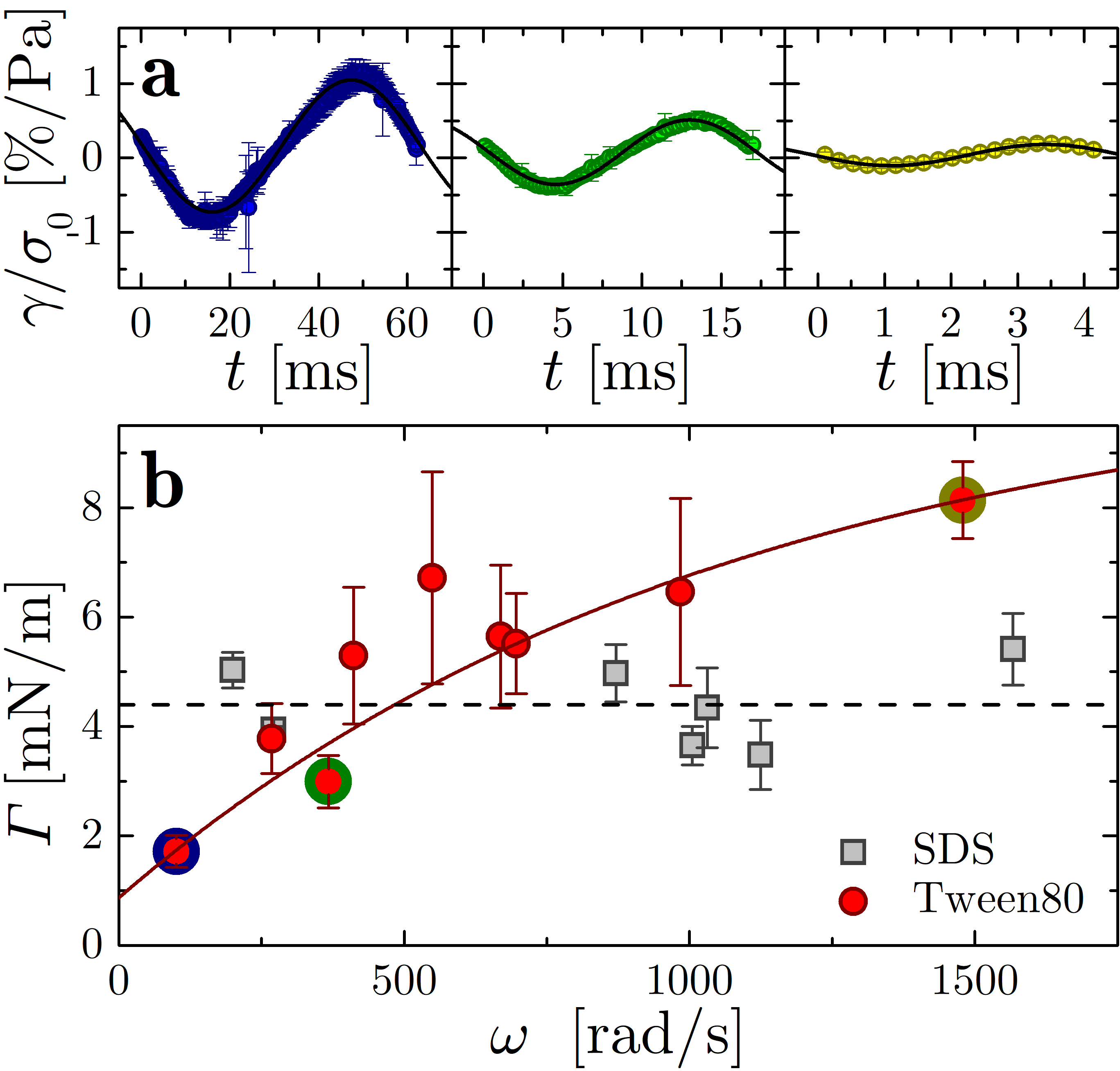}
\caption{\label{fig.4Freq}\textbf{Dynamic surface tension} (a) Symbols: time-dependent deformation profiles measured for $\omega=100$, 350 and 1500~rad/s (left to right). Lines: sinusoidal fits. (b) Symbols: surface tension for oil droplets stabilized by SDS (gray squares), and Tween80 (red circles). Error bars: data dispersion over many droplets. Red line: exponential fit.}
\end{figure}

Surprisingly, we observe that the droplets are stiffer when probed at higher $\omega$, as they deform less for under a given stress amplitude, as shown in Fig.~\ref{fig.4Freq}a. 
To quantify this effect, we extract the frequency-dependent surface tension, $\Gamma(\omega)$, and we find that it increases with $\omega$. We find that data are well fit by the empirical model $\Gamma(\omega)=\Gamma_0+(\Gamma_\infty-\Gamma_0)\left[1-\exp(-\omega \tau)\right]$, where $\tau\approx1$~ms is the relaxation time of the droplet interface, as shown by the red circles and solid line in Fig.~\ref{fig.4Freq}b. 
This empirical model assumes that the adsorption and equilibration of surfactant molecules at the interface occurs in a time $\tau$: deformations at frequencies \(\omega \gg \tau^{-1}\) result in a poorly-equilibrated interface, with larger $\Gamma$. 
To test this hypothesis, we change the surfactant, replacing Tween80 with sodium dodecyl sulfate (SDS): being smaller and more water soluble, we expect $\tau$ to decrease. Indeed, for SDS-stabilized droplets, we find a frequency-independent $\Gamma=4.4\pm0.7$~mN/m, as shown by the gray squares in Fig.~\ref{fig.4Freq}b. This suggests that SDS-stabilized droplets have a much faster relaxation time, below 0.1~s, corroborating the physical interpretation of $\tau$ as a surfactant equilibration time. 
This result demonstrates that Rheofluidics can be used to extend conventional tensiometry to the high-frequency regime, where $\Gamma(\omega)$ reveals the unexpected frequency-dependent behavior of well-known surfactants at liquid-liquid interfaces. 

\begin{figure}[ht]
\includegraphics[width=\columnwidth]{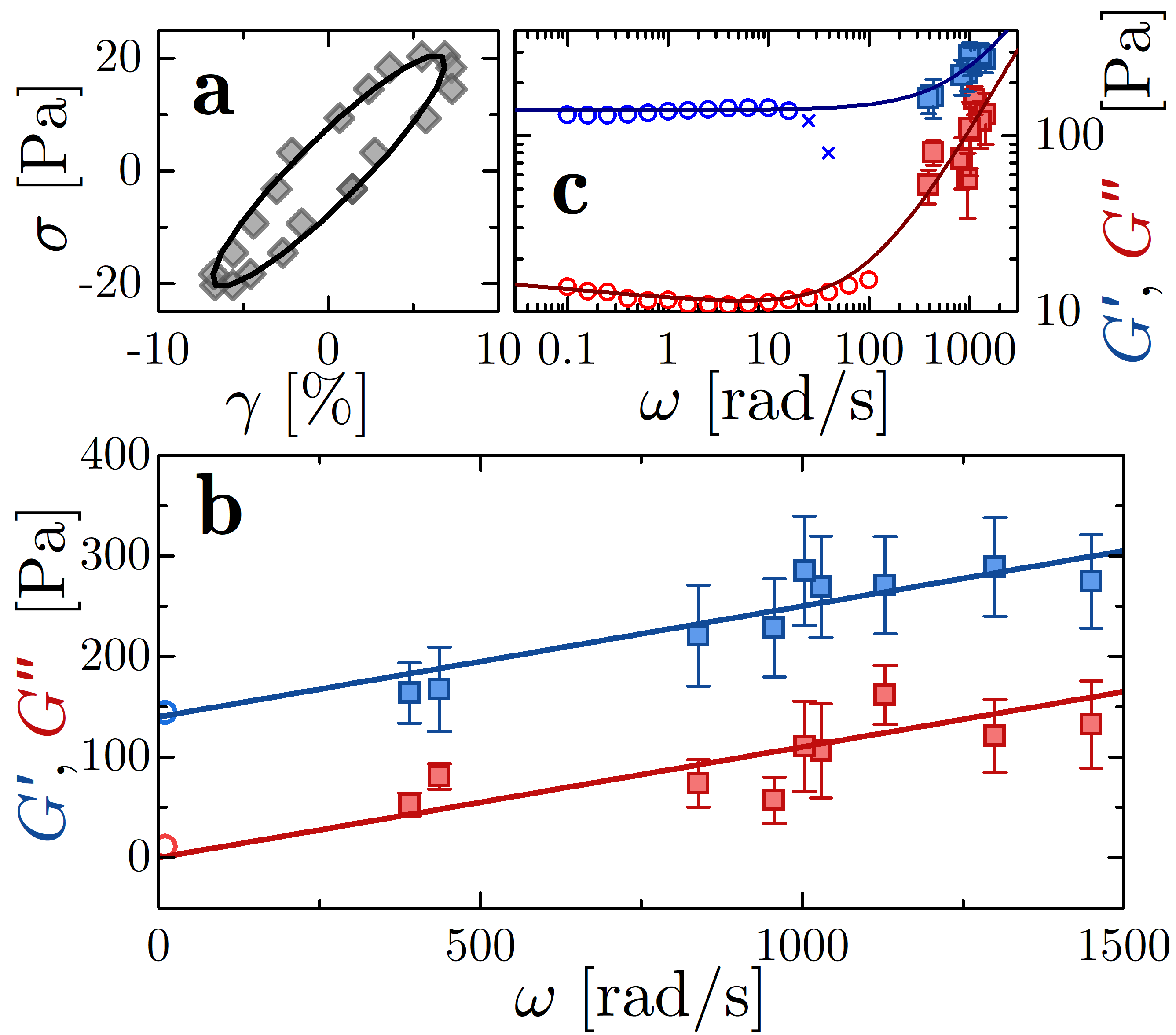}
\caption{\label{fig.5beads}\textbf{Hydrogel viscoelasticity} (a) Gray diamonds: Lissajous figure for $\omega=10^3$~rad/s. Black line: fitted $\sigma(\gamma)$. (b) Squares: storage (blue) and loss (red) moduli measured by Rheofluidics. Solid lines are linear fits, with $G_0=140$~Pa and $\tau=0.8$~ms.
Open circles: linear viscoelastic moduli measured by shear rheology in bulk hydrogels at $\omega=10$~rad/s. (c) Open circles: $G^\prime$, $G^{\prime\prime}$ for a bulk hydrogel measured by shear rheology; Full symbols: same data as panel b}
\end{figure}

Rheofluidics can also measure the mechanical properties of solid particles, such as hydrogel beads, whose mechanical properties are dominated by the viscoelasticity of the gel. We illustrate this by investigating alginate hydrogel particles, 35~$\mu$m in diameter, crosslinked by calcium ions. To increase their optical contrast, we dye them with metylene blue, and we flow them in the Rheofluidic channel, adjusting the flow rate to obtain a stress amplitude $\sigma_0=20$ Pa. To highlight the viscoelastic response of the beads, we plot stress as a function of strain, obtaining the elliptical Lissajous figure shown in Fig.~\ref{fig.5beads}a. 
We analyze it to extract the viscoelastic moduli of the bead, and we find that both $G^\prime$ and $G^{\prime\prime}$ increase linearly with $\omega$, as captured by the empirical model: $G^\prime=G_0+\eta\omega$ and $G^{\prime\prime}=\eta\omega$, where $G_0=140$~Pa is the low-frequency plateau of the bulk elastic modulus of the hydrogel, and $\eta=100~\mPas$ is an effective viscosity. We find that this value of viscosity is close to that of the 1:3 water:glycerol mixture used a continuous phase, $\eta_s=58~\mPas$. A similar frequency dependence has been observed in the high-frequency response of many soft glassy materials~\cite{prasadRidealLectureUniversal2003}. 
To confirm that Rheofluidics can be used to measure the high-frequency rheology of soft solids, we crosslink a bulk alginate hydrogel in a rheometer, and we measure its linear viscoelastic moduli. We find $G^\prime\sim 140$~Pa, almost constant up to $\omega\approx10$~rad/s, above which bulk rheology becomes unreliable due to inertia~\cite{ewoldt_experimental_2015}, while $G^{\prime\prime}$ is about one decade smaller, as shown by circles in Fig.~\ref{fig.5beads}c. 
Bulk rheology and Rheofluidics provide a consistent physical picture, which demonstrates the power of Rheofluidics to measure the rheology of viscoelastic solids, accessing much larger frequencies compared to rheometers due to the low inertia. 

\begin{figure}[ht]
\includegraphics[width=\columnwidth]{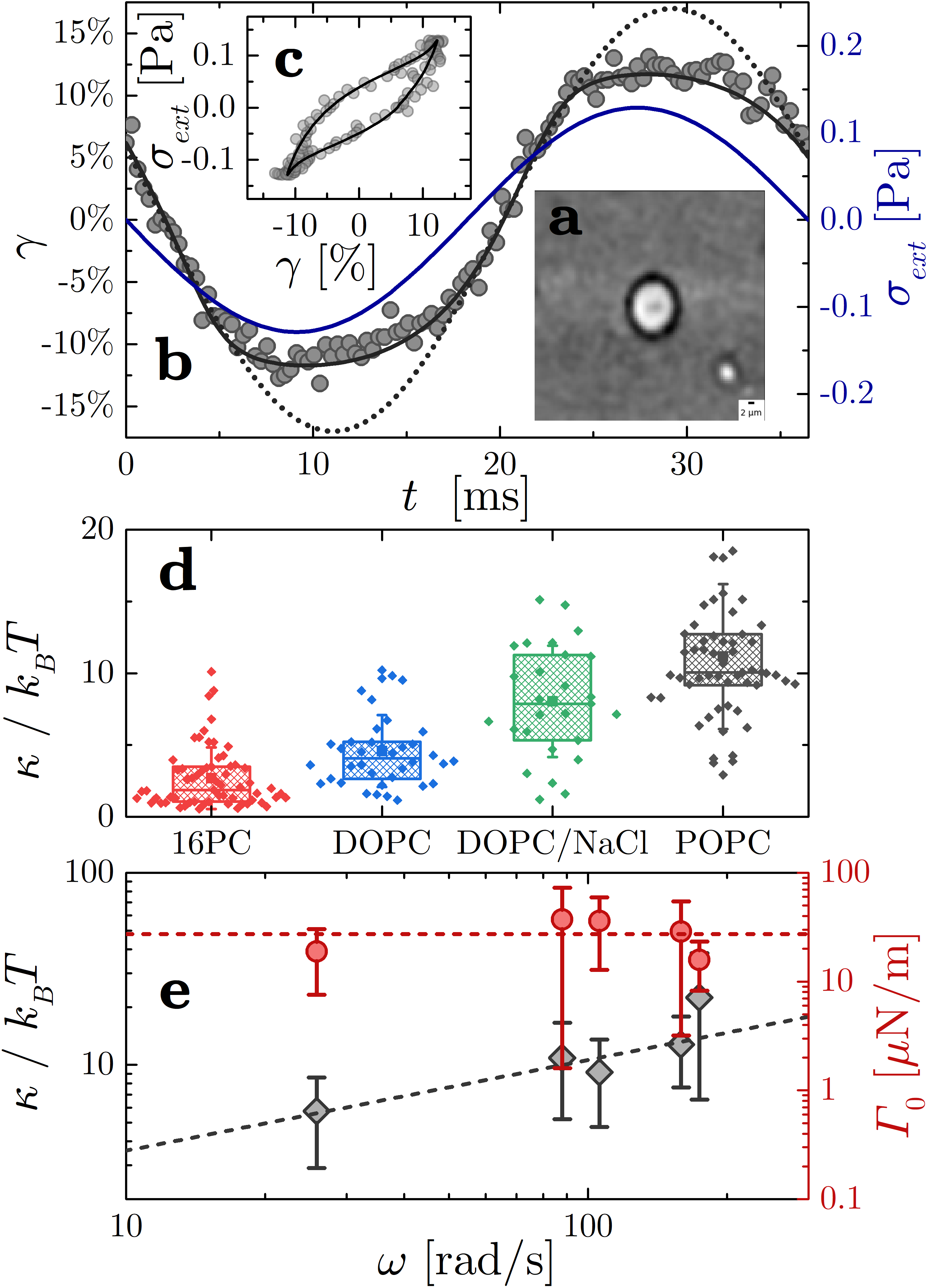}
\caption{\label{fig.6vesicles}\textbf{Lipid vesicles} (a) image of a POPC vesicle. (b) blue line, right axis: extensional stress profile. Symbols, right axis: vesicle deformation. Black solid line: nonlinear model fitted to the data. Dashed line: model without strain stiffening, highlighting deviations from the sinusoidal deformation profile. (c) Lissajous figure constructed with data in panel a. (d) Bending stiffness measured at $\omega=100$rad/s for different vesicle compositions (e) Frequency-dependent bending stiffness (gray symbols, left axis) and membrane tension (red symbols, right axis) for POPC. Error bars represent data dispersion. Dashed lines are a guide to the eye.}
\end{figure}

The fine control of the applied stress, down to a few mPa, enables the use of Rheofluidics to measure ultrasoft objects, and even beyond the linear regime. To demonstrate this, we measure Giant Unilamellar Vesicles (GUVs). 
Static measurements of GUV deformability are obtained using micropipette aspiration~\cite{evansEntropydrivenTensionBending1990b}, and are used to extract the bending stiffness, $\kappa$, of the membrane. 
To measure GUVs with Rheofluidics, we form 1-palmitoyl-2-oleoyl-glycero-3-phosphocholine (POPC) vesicles in a water-glucose mixture, and then disperse them in a water-sucrose mixture of same osmolarity and viscosity $\eta_s=1.3~\mPas$, but different refractive index, enabling  easier visualization of the flowing GUVs, as shown in Fig.~\ref{fig.5beads}a. 
GUVs are highly deformable, thus their measurement is delicate, and requires a fine control of the applied stress~\cite{morshedMechanicalCharacterizationVesicles2020}. Rheofluidics is an ideal technique to apply small stresses: to this end, we reduce the volumetric flow rate to obtain a stress amplitude $\sigma_0=0.13$~Pa and frequency $\omega=170$~rad/s, shown by the blue line in Fig.~\ref{fig.6vesicles}b. 
We observe that $\gamma(t)$ is not purely sinusoidal, but instead nearly saturates around 10\% deformation, as shown by gray symbols in Fig.~\ref{fig.6vesicles}b. This saturation is indicative of nonlinear strain stiffening, as better visualized by the stress-strain curve, which exhibits steeper sections for $\gamma \approx 10\%$, as shown in Fig.~\ref{fig.6vesicles}c. 
This reflects the straightening of small-scale membrane fluctuations, which become increasingly constrained as the vesicle is deformed at constant volume, entailing an exponential increase of the membrane tension \cite{evansEntropydrivenTensionBending1990b}. This effect can be modeled by introducing an effective bending stiffness, $\kappa$, describing the strain dependence of the membrane tension: $\Gamma=\Gamma_0\exp(\kappa\alpha)$, where $\alpha\propto\gamma^2$ is the areal strain~\cite{SM}. As a result of the increasing $\Gamma(\gamma)$, vesicles get stiffer as they deform: thus, a sinusoidal stress results in a non-sinusoidal strain. We find that this model fits the measured $\gamma(t)$ very well, allowing us to measure $\Gamma_0$ and $\kappa$ for each vesicle. We collect data for $\sim 60$ vesicles with diameters between 8 and 20~$\mu$m, and we find no measurable size dependence. Averaging measurements for all droplets, we obtain $\Gamma_0= 30~\mu$N/m and $\kappa=11k_BT$, in fair agreement with literature data \cite{dimova_recent_2014,solmaz_optical_2013}.

We repeat the measurement replacing POPC with fully unsaturated lipid molecules such as 1,2-dioleoyl-sn-glycero-3-phosphocholine (DOPC) and 1,2-dipalmitoleoyl-sn-glycero-3-phosphocholine (16PC), which loosely pack at the interface, resulting in more flexible membranes~\cite{de_mel_influence_2020}. In both cases, $\kappa$ is 2-5 times smaller at $\omega=100$~rad/s. To corroborate this result, we improve lipid packing by adding 100mM NaCl to reduce the electrostatic repulsion between head groups. We find that this indeed results in stiffer DOPC vesicles, as shown in Fig.~\ref{fig.6vesicles}d. This composition dependence, usually characterized using time-consuming techniques such as micropipette aspiration, is here captured by analyzing four videos lasting a few seconds. 
Finally, we measure the frequency-dependent mechanical properties of POPC vesicles over an unexplored range of high frequencies. We find that $\Gamma_0$ is independent of $\omega$, confirming its physical interpretation as the membrane tension of the undeformed vesicle. 
By contrast, our data suggest that in this high frequency regime $\kappa$ increases with $\omega$, as shown in Fig.~\ref{fig.6vesicles}e. 
This behavior may reflect a reduced contribution from a portion of the spectrum of thermally-excited modes of the vesicle fluctuations; the high-frequency Rheofluidic measurements limit the contributions of the low-frequency modes, thereby increasing $\kappa(\omega)$.
These surprising results call for more detailed experimental validation, but they highlight the power of Rheofluidics to characterize high frequency mechanical behavior in lipid vesicles.

These results demonstrate that Rheofluidics enables measurement of the frequency-dependent moduli of individual soft matter objects, in a regime of high frequencies that probe fast physical processes, including adsorption and equilibration of surfactants at fluid interfaces and fluctuations of lipid membranes. For particles such as hydrogel beads, it extends rheological measurements to much higher frequencies, which cannot be measured with rheometers due to their inertia.
Moreover, Rheofluidics is also well-suited for analyzing heterogeneous and time-evolving samples, making it a promising technique to address challenges in fields spanning fundamental physics, interfacial science, biophysics, and cell biology.

\textit{Data availability}—The data that support the findings of this article are openly available~\cite{zenodo}, as well as the python code used for data analysis~\cite{Rheoflupy}; embargo periods may apply

\bibliography{apssamp}

\end{document}